\documentclass[twocolumn,showpacs,preprintnumbers,amsmath,amssymb]{revtex4}

\usepackage{graphicx}
\usepackage{dcolumn}
\usepackage{bm}
\usepackage[left]{lineno}
\usepackage[usenames]{color}

\begin{document}

\title{Strain-induced pseudo-magnetic field for novel graphene electronics}

\author{Tony Low}
\email{tonyaslow@gmail.com }
\affiliation{$^*$Network for Computational Nanoelectronics, Hall for Discovery Learning Research \\
Purdue University, West Lafayette, IN47907-1791 Indiana, US\\
$^\dagger$Instituto de Ciencia de Materiales de Madrid, CSIC, Sor
Juana In\'es de la Cruz 3, E28049 Madrid, Spain}

\author{F. Guinea}
\email{paco.guinea@icmm.csic.es}
\affiliation{$^*$Network for Computational Nanoelectronics, Hall for Discovery Learning Research \\
Purdue University, West Lafayette, IN47907-1791 Indiana, US\\
$^\dagger$Instituto de Ciencia de Materiales de Madrid, CSIC, Sor
Juana In\'es de la Cruz 3, E28049 Madrid, Spain}

\begin{abstract}
Particular strain geometry in graphene could leads to a uniform
pseudo-magnetic field of order $10T$ and might open up interesting 
applications in graphene nano-electronics. Through quantum transport 
calculations of realistic strained graphene flakes of sizes of $100nm$, 
we examine possible means of exploiting this effect for practical electronics and 
valleytronics devices. First, we found that elastic backscattering at rough edges
leads to the formation of well defined transport gaps of order $100meV$
under moderate maximum strain of $10\%$. Second, the application of a real magnetic field 
induced a separation, in space and energy, of the states arising from different valleys, leading to
a way of inducing bulk valley polarization which is insensitive to short
range scattering.
\end{abstract}

\maketitle

\newpage

{\em Introduction.} The isolation of single layer graphene, and the
possibility of controlling the electronic
carrier density\cite{Netal04,Netal05,Netal05b} has led to a large research
effort, because of the novel fundamental features exhibited by
graphene and also its possible applications\cite{NGPNG09,AGSci09}.
Graphene is a two dimensional membrane whose electronic properties
can be controlled  by applying a gate voltage. It can be shown that
the elastic deformations of the membrane modify the electronic
properties, as they play the role of an effective gauge
field\cite{SA02b,M07,NGPNG09}, opening a new way of tailoring the
electronic properties. Strains of at least 10\% can be induced in
graphene without damaging appreciably its structure\cite{Hone08}.
Suspended graphene samples show long range deformations on scales of
hundreds of nanometers\cite{Betal07,Betal08,Betal09}, and strains
can be induced in graphene samples by different
techniques\cite{Metal09p,Hetal09,Petal09}.

Strains can be expected to arise naturally in suspended
samples\cite{FGK08}. It has been shown that a uniformly varying
strain leads to a gauge potential which generates an effective
constant magnetic field\cite{GKG10,GGKN10} (time reversal symmetry
implies that the field has opposite signs in the two valleys). The
field due to strains can interfere in many ways with real magnetic
fields\cite{PSLFG09}. Other schemes to use strains\cite{PN09} to
tune the electronic properties have also been proposed. Strains can
possibly be manipulated efficiently in samples with good adhesion to
the substrate, such as graphene layers grown epitaxially on SiC.
Hence, the proposed effects in this paper could be tested experimentally.

\begin{figure}[htps]
\centering \scalebox{0.42}[0.42]{\includegraphics*[viewport=110 40 700
580]{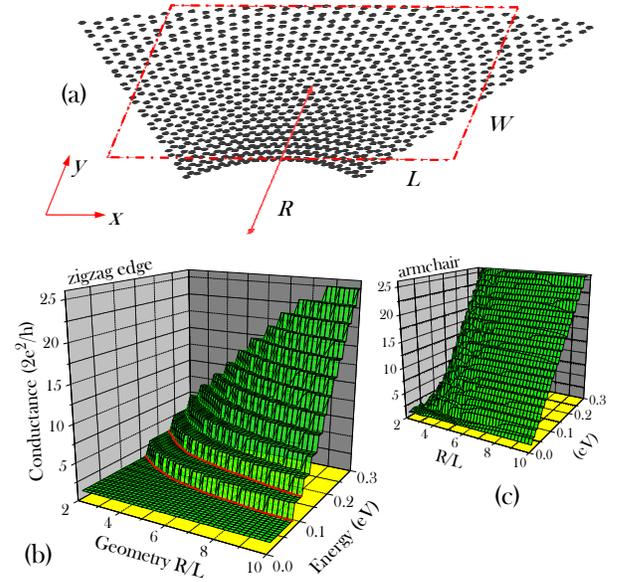}}\caption[fig]{(a) Sketch of an example strain geometry with
a maximum strain of $50\%$. Note that the actual size of the flake used in this paper is much larger, i.e. $L=W=100$nm, and with much smaller maximum strain (see text).
Conductance as a function of
Fermi energy and  strained geometry characterized by $R/W$ for (b)
zigzag and (c) armchair edge ribbons, in the absence of real
magnetic field and edge disorder. The dimension of the graphene
flake is $L=W=100$nm.}
 \label{FIG1}
\end{figure}

As discussed in\cite{GGKN10}, an effective constant magnetic field
arises from strains varying at a constant rate.  We
discuss here the electronic properties of graphene flakes under the
combined effect of strains, magnetic fields and disorder. We use the
setup proposed in\cite{GGKN10} as a way of inducing an almost
constant effective magnetic field. The model and method of
calculation are discussed in the next section. Then, we present the
main results. The last section
discusses the main conclusions of our work.

{\em The model.}
 A strain distribution, $u_{ij}$, leads to the effective gauge field\cite{SA02b,M07,GKG10},
\begin{eqnarray}
\tilde{\bf A}=\frac{c\beta}{a_{0}}\left(
\begin{array}{c}
u_{xx}-u_{yy}\\
-2u_{xy}
\end{array}\right)
\label{pseudogauge}
\end{eqnarray}
where $\beta\approx\partial \mbox{log}(t)/\partial
\mbox{log}(a)|_{a=a_{0}} \approx 2-3$,  $a_{0} \approx 1.4$ \AA \, is
the lattice constant, $t \approx 3$eV is the nearest neighbor
coupling energy and $c$ is a dimensionless constant of order unity.
The main axes of the graphene lattice in
Eq.~\ref{pseudogauge} has been chosen so that the $x$ axis coincides with a zigzag
direction.  We assume that the distortions have been induced by
bending a flake in the way discussed in \cite{GGKN10}
\begin{eqnarray}
(u_{x},u_{y})=\left(\frac{xy}{R},-\frac{x^{2}}{2R}\right)
\label{arcdeform}
\end{eqnarray}
where $R$ is the bending radius of the deformation applied to the
flake, see Fig.~\ref{FIG1}a. This would
translates to a pseudomagnetic field strength of $B_{s}=c\beta/a_{0}R\equiv \Omega W/R$,
where $\Omega$ is to be determined numerically.

\begin{figure}[t]
\centering \scalebox{0.5}[0.5]{\includegraphics*[viewport=55 230
650 590]{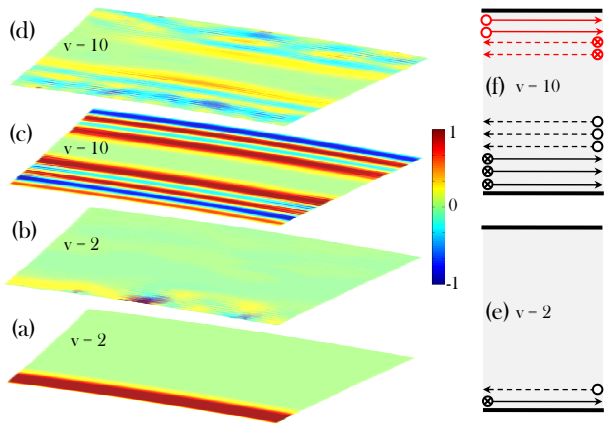}} \caption{\footnotesize  The non-equilibrium
current density
$\mbox{sign}(\mbox{j}_{x})\times\left|\bold{j}(x,y)\right|$
intensity plot for a $100\times 100nm$ graphene flake under a strain
of $R/W=5$ (equivalent to $B_{s}\approx 9$T) at $\epsilon_{F}$
corresponding to filling factor $\nu=2$ (a) without edge disorder and
(b) with edge disorder. Similar plots in (c-d) but for filling factor $\nu=10$
instead. The color scale indicated is normalized with respect to 
$max(\left|\bold{j}(x,y)\right|)$. Red/blue color indicates 
current flowing to the right/left of the device.
(e) and (f) illustrates the counter-propagating edge states
at $v=2$ and $v=10$ respectively.  } \label{FIG2}
\end{figure}

To compute the transport properties, we adopt a nearest  neighbor
hopping between $\pi$ orbitals in the honeycomb lattice
\cite{wallace47},
\begin{eqnarray}
{\cal H}=\sum_{ij}\left[t_{0} \left( 1 +\beta \frac{\delta
d_{ij}}{a_0} \right)
\right]\mbox{exp}\left(\frac{e}{\hbar}\int_{i}^{j}\bold{A}\cdot
d\bold{l}\right)a_{i}^{\dagger}a_{j} \label{tightbinding}
\end{eqnarray}
where $\delta d_{ij}$ is the  change in bond length, and $\bold{A}$
is the real gauge field, whose effect is incorporated as a Peierls
phase. We assumed $t_{0}=3$eV and $\beta=2$. Semi-infinite leads
are assumed for the left and right boundaries. The numerical methods
that we had employed to compute the transmission function $T$ is the
recursive green function\cite{datta97,nonoyama98,anantram08} and
the renormalization method\cite{grosso89}. The Landauer formula
then gives us the zero temperature conductance i.e. $2e^{2}/h\times
T$\cite{landauer70}. Details of the implementation had
already been documented elsewhere\cite{low09prb1,low09prb2}.

We consider square graphene flakes with dimensions $W=L=100$nm where
the origin for the deformation in Eq. \ref{arcdeform} is the center
of the flake. The maximum strain exerted would then be along the two
edges, which is given by $\mbox{max}(u_{xx})\approx W/2R$.

{\em Results.} We show in Fig.\ref{FIG1}b and c  the computed
conductance as a function of $R/W$ and energy for zigzag and
armchair edge ribbons respectively. Clean quantum Hall plateaus are
observed for the zigzag ribbons with filling factors given by
$v=2,6,10\ldots=4n+2$, exactly mimicking the conventional quantum
Hall case. Knowing that the first excited Landau energy
$\epsilon_{1}=v_{F}\sqrt{2\hbar Be}$ in the conventional quantum
Hall case, we can estimate numerically the effective magnetic
field induced by a deformation characterized by the ratio $W/R$ as
 for $B_{s}= \Omega W/R$ with $\Omega\approx 45$T, in good agreement with the estimates
 in~\cite{GKG10,GGKN10}.

\begin{figure}[t]
\centering \scalebox{0.35}[0.35]{\includegraphics*[viewport=5 20 650
620]{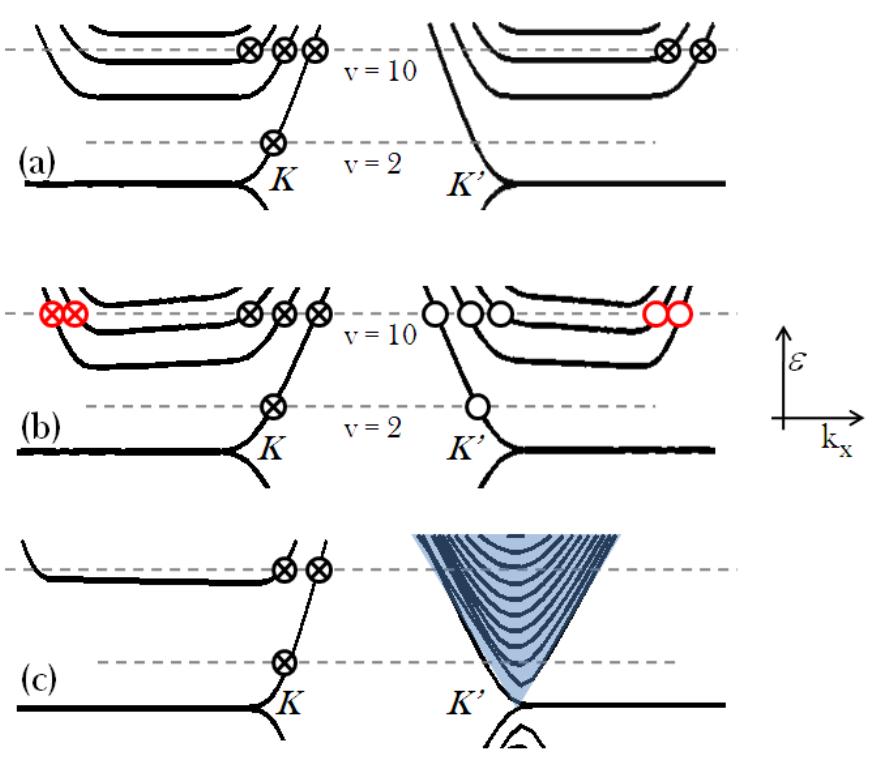}} \caption{\footnotesize  Plot of typical energy dispersion as a
function of momentum along the transport direction for the case of
(a) real magnetic field $B=9T$, (b) pseudomagnetic field $B_{s}=9T$
and (c) $B_{s}=B=9T$. The symbols $\otimes$ and $\mbox{O}$ denotes 
the sign of the field, while red/black color denotes whether the chiral 
edge states are
propagating along the top/bottom edge. The highlighted bulk states
in (c) are non-chiral i.e. it feels an effective zero magnetic field.  } \label{FIG3}
\end{figure}

The effects on the electronic states of  a uniaxial strain, $u_{xx}
( x )$, like the one studied here, depend strongly on the
orientation of the lattice with respect to the direction of the
strains. When the $y$ axis coincides with an armchair direction, a
gauge field along the $x$ direction is generated, $A_x ( y )$, and
an effective magnetic field ensues (see for example simpler
case of a one-dimensional ripple \cite{guinea081dripple}).
On the other hand, when the $y$
axis is along a zigzag direction, the gauge field due to the
uniaxial strain can be written as $A_y ( y )$. This
gauge field does not induce an effective magnetic field, and leaves
the electronic spectrum unchanged. The results in Fig.~\ref{FIG1}
are in agreement with this analysis.
 Hereafter, we shall only consider zigzag ribbons
with strain geometry $R/W=5$, which yields $B_{s}\approx 9$T.

Unlike the case in the Quantum Hall effect, the edge states are not
protected by time reversal symmetry, and they can be affected by
elastic backscattering. Fig.~\ref{FIG2} a and c plots the
non-equilibrium current density at Fermi energy corresponding to
$\nu=2$ and $\nu=10$. For $\nu=2$, strains induce two edge modes,
which propagate in opposite directions. Time reversal symmetry
implies that these two modes are localized at the same edge, in
agreement with the numerical results. In general, the compressive 
strained edge would acquire two more modes than the other edge. 
The zigzag boundary condition 
used here do not mix the $K$ and $K'$
valleys, leading to a clear distinction of the edge modes.

\begin{figure}[t]
\centering \scalebox{0.53}[0.53]{\includegraphics*[viewport=130 240
550 530]{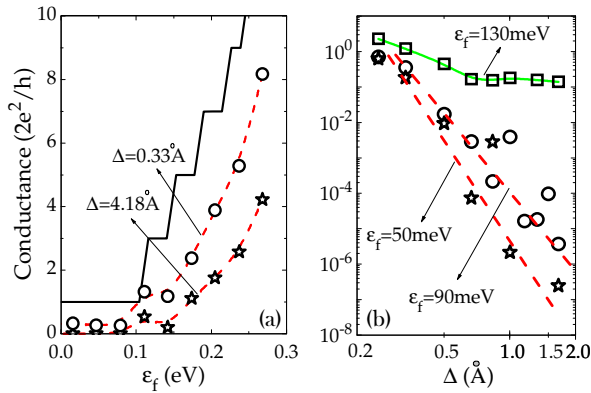}} \caption{\footnotesize  Zero temperature
conductance as a function of (a)  Fermi energy $\epsilon_{f}$ and
(b) edge disorder $\Delta$, for different values of $\Delta$ and
$\epsilon_{f}$ respectively. Zigzag ribbon is assumed and the strain
geometry employed corresponds to $W/R=5$. The solid line
for (a) is for the case of zero disorder. } \label{FIG4}
\end{figure}

The zigzag boundary condition (where a given edge can be characterized
by a majority sublattice) is generic for graphene
edges\cite{AB08}, except the armchair one. The edge modes induced by
strains have a characteristic width of order of the effective
magnetic length associated to the strain field, $\ell_{s} \approx
\sqrt{( a_{0} L)/(\beta \bar{u})}$, where $L$ is a typical length which
describes the variation of the strain, and $\bar{u}$ is the average
strain. Typically $\ell_{s} \gg a_0$, and the continuum zigzag
boundary conditions discussed in\cite{AB08} describe well the
numerical results.

The physical origin of the edge currents is explained in Fig.
\ref{FIG3}a and b, which plots the corresponding energy dispersion
along the transport direction for the case with a real and
pseudomagnetic field respectively. For the
latter, one makes the observation that for a given current
direction, the edge states on the two edges are valley polarized
i.e. quantum valley Hall effect. This effect is analogous to the
quantum spin Hall effect\cite{kane05}. In both cases, the net
pseudo-gauge field of the system is zero, but finite and opposite
for each spin/valley. However, in this case, short-range scattering
would couples the valleys. Since the counter propagating edge
states residing along a particular edge belongs to different
valleys, intervalley processes lead to backscattering. In the
presence of edge disorder, substantial backscattering can occurs and
Anderson localization spots can be observed (see Fig. \ref{FIG2}b).
A more quantitative evaluation of the impact of edge roughness
follows.

\begin{figure}[t]
\scalebox{0.3}[0.3]{\includegraphics*[viewport=30 10 790
900]{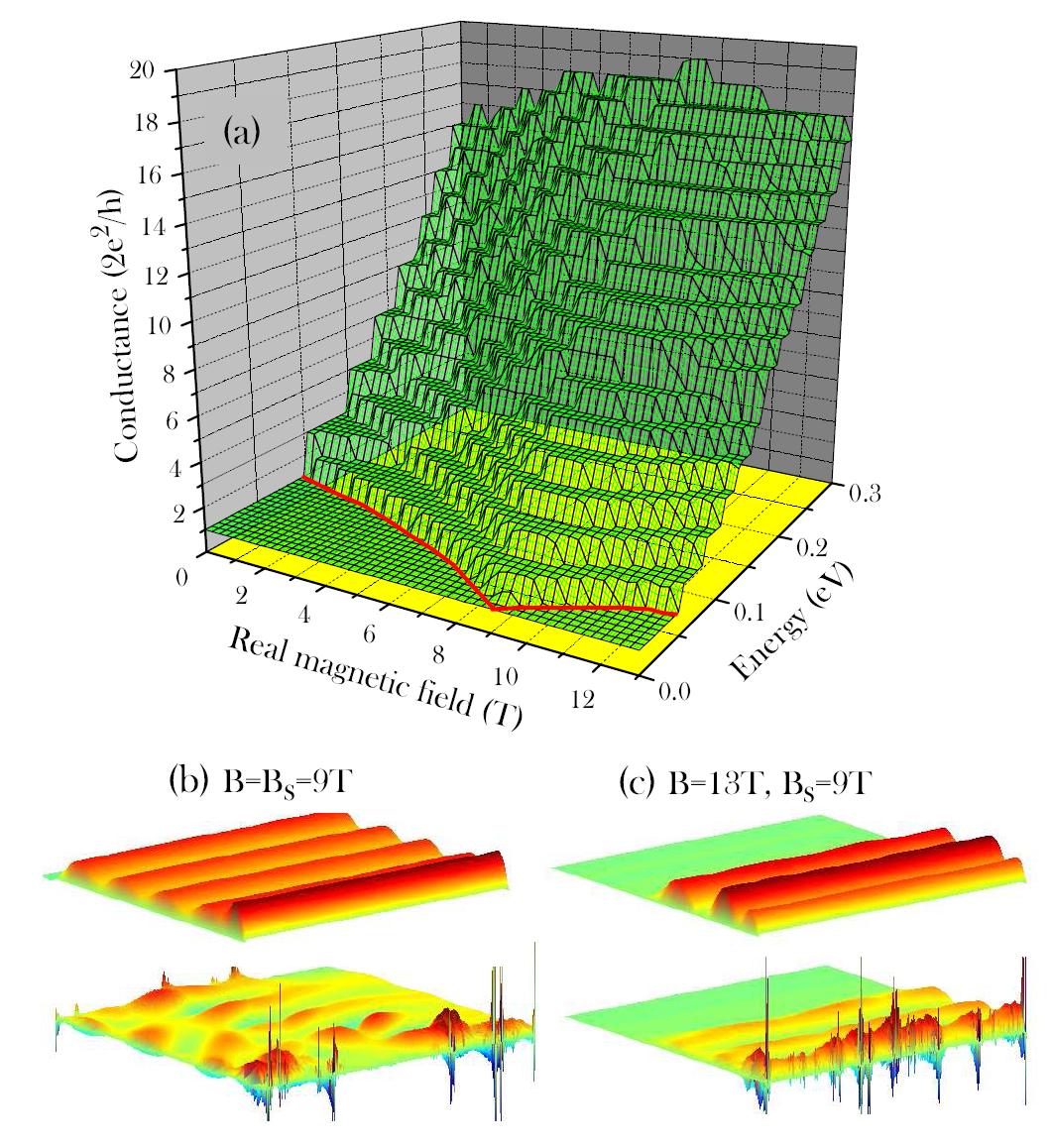}} \caption{\footnotesize  (a) shows the conductance as a function of
real magnetic field and Fermi energy, calculated for non-disordered
zigzag ribbon with a strain geometry corresponding to $W/R=5$, which
is equivalent to $B_{s}\approx 9T$. (b) plots the current density
at $\epsilon_{f}=0.1eV$ for the condition $B=B_{s}=9T$, for perfect edge (top)
and disorder edges (bottom). Similar plots for (c), except now for the
condition of $B=13T$ and $B_{s}=9T$. The color scheme employed is the same
as Fig. \ref{FIG2}. } \label{FIG5}
\end{figure}

The edge roughness is generated using a procedure outlined in\cite{low09prb2}.
The roughness morphologies generated were assumed
to have an autocorrelation length of $2$nm and a root-mean-square
roughness $\Delta$, where $0.2$\AA$\leq\Delta\leq 5$\AA \, is used
as the disorder parameter in this work. Fig. \ref{FIG4}a plots the
energy resolved conductance for various values of $\Delta$. Disorder
at the edges opens a transport gap $\epsilon_{g}$, and the flake
shows insulating behavior near the neutrality point. When the
disorder increases, $\epsilon_{g}$ saturates to a value of the order
of two times the energy of the first effective Landau level,
modulated by the strain via $\epsilon_{g}\propto \sqrt{W/R}$. In the
nanoribbon counterpart, it was found to behave as
$\epsilon_{g}\propto 1/W$\cite{mucciolo09}. The pseudomagnetic
length is given by $\ell_{s} \propto \sqrt{R/W}$, which
approximates the spatial extent of the ground state Landau
wavefunction from the edges, can be of order of $10nm$ at moderate
strain of $W/R=5$. The estimates in the Supplementary Information
in\cite{GKG10} suggest a localization length for the edge modes, due
to intervalley scattering of order $\xi \approx ( \ell_{s} / a_{0} )^2
n_v^{-1}$, where $n_v$ is the one dimensional density of vacancies
at the edges. For a rough edge on atomic scales, $n_v \sim
10$nm$^{-1}$, so that $\xi \sim W$, and localization can be expected
for the strains and dimensions used here.

Next, we examine the interplay between real and pseudomagnetic
field. Their combined  contribution is such that the valleys feel an
effective field of $\tilde{\bold{A}}\pm\bold{A}$. Fig. \ref{FIG5}a
shows the conductance for varying $B$ field but at $B_{s}\approx
9T$. Splitting of the conductance plateaus steps from $4n+2$ into
$2n+2$ can be observed, with decreasing plateau width for $n=0$. An
interesting situation arises when $B=B_{s}$, where Fig. \ref{FIG3}c
illustrates the corresponding energy dispersion. The cancellation of
the field for $\bold{K}'$ valley leads to a recovery of the bulk
band dispersion. Hence, the conductance goes as $\approx
c\epsilon_{f}+2$, where the latter contribution comes from the
$\bold{K}$ valley. This provides a feasible avenue for producing
valley polarized electrons, since it does not depend on good quality
at the edges like in previous proposal\cite{rycerz07}.
\textcolor{black}{We note that in general,
these bulk states are not protected by time reversal symmetry.
Fig. \ref{FIG5}b plots the current density at the condition $B=B_{s}=9T$,
with/without edge disorder. Their respective conductances (in $2e^{2}/h$) are
$5$ and $3.3$ respectively. At larger magnetic field, say $B=13T>B_{s}=9T$,
edge states emerge and now the conductance remains at $3$ with/without edge disorder.
Fig. \ref{FIG5}c shows the respective current density plots.}

{\em Conclusions.} We have analyzed in detail the electronic
properties of graphene flakes under the combined effect of strains,
magnetic fields and disorder. We analyze numerically the electronic
spectrum of flakes of realistic sizes, $\sim 100$nm with strain
distributions that generate a constant effective magnetic field.
Strains induce gaps in the bulk spectrum and propagating modes along
the edges. The edge modes are valley polarized, and strongly
suppressed at armchair edges, where the boundary condition strongly
mixes the different valleys \cite{brey06}. The two lowest edge modes
are localized at the same edge. In general, the number of edge modes
differs by two between the two edges.

Possible device applications are addressed. First,
disorder leads to backscattering, which is more significant for the
lowest edge (see Fig. \ref{FIG4}b). We find a clear transport gap near the Dirac energy,
leading to new ways of developing graphene transistors.
Second, the interference between real magnetic fields and the gauge field
due to strains lead to the  separation in space and in energy of the
states from the two valleys. Because of this separation, the valley
polarization achieved in this way is not much affected by
intervalley scattering, opening a way of obtaining valley polarized
bulk currents.

{\em Acknowledgements.} We are grateful to A. K. Geim, M. I.
Katsnelson, K. S. Novoselov and M. Lundstrom for useful discussions
and Network of Computational Nanoelectronics for computational support.
F. G. acknowledges funding from MICINN (Spain), through grants
FIS2008-00124 and CONSOLIDER CSD2007-00010. T. L. acknowledges
funding from the Institute for Nanoelectronics
Discovery and Exploration.


\end{document}